\documentclass[useAMS,usenatbib]{mn2e} 
\usepackage{psfig} 
 
\def\cbeta{$c_{\beta}$} 
\def\kms{\relax \ifmmode {\,\rm km\,s}^{-1}\else \,km\,s$^{-1}$\fi}

\def\mincir{\ \raise-2.truept\hbox{\rlap{\hbox{$\sim$}}\raise5.truept 
    \hbox{$<$}\ }} 
\def\magcir{\ \raise-2.truept\hbox{\rlap{\hbox{$\sim$}}\raise5.truept 
    \hbox{$>$}\ }}

\def\nii{[N {\sc ii}]} 
\def\hi{H {\sc i}} 
\def\hii{H {\sc ii}} 
\def\sii{[S {\sc ii}]} 
\def\siii{[S {\sc iii}]} 
\def\ariii{[Ar {\sc iii}]} 
\def\oii{[O {\sc ii}]} 
 
\def\heii{He{\sc ii}} 
\def\hei{He{\sc i}} 
\def\oiii{[O {\sc iii}]} 
\def\neiii{[Ne {\sc iii}]} 
 
\def\ha{H$\alpha$} 
\def\hb{H$\beta$} 
\def\hd{H$\delta$} 
\def\hg{H$\gamma$} 
 
\def\te{$T_e$} 
 
\def\ne{$N_e$} 
\def\nh{$N_H$} 
\def\di{dIrs} 
\def\ds{dSphs} 
 
\title[Abundances from PNe in NGC~147]{The chemical content of nearby 
galaxies from planetary nebulae: NGC~147\thanks{Based on observations 
obtained at the Gemini Observatory, which is operated by the 
Association of Universities for Research in Astronomy, Inc., under a 
cooperative agreement with the NSF on behalf of the Gemini 
partnership.}} 
 
\author[Gon\c calves et al.]{D. R. Gon\c calves$^{1}$\thanks{E-mail: 
denise@astro.iag.usp.br}; 
L. Magrini$^{2}$; P. Leisy$^{3,4}$ and R.L.M. Corradi$^{3,4}$ 
\\ 
$^{1}$IAG, Universidade de S\~ao Paulo, Rua do Mat\~ao 1226, 05508-900 S\~ao Paulo, 
Brazil\\ 
$^{2}$INAF, Osservatorio Astrofisico di Arcetri, Largo E. Fermi, 5. I-50125, Firenze, Italy\\ 
$^{3}$Isaac Newton Group of Telescopes, Apartado de Correos 321, E-38700 
Sta. Cruz de La Palma, Spain\\ 
$^{4}$ Instituto de Astrof\'{\i}sica de Canarias, E-38205 La Laguna, 
Tenerife, Spain} 
 
\begin{document} 
 
\date{Accepted ?. Received ?; in original form ?} 
 
\pagerange{\pageref{firstpage}--\pageref{lastpage}} \pubyear{2006} 
 
\maketitle 
 
\label{firstpage} 
 
\begin{abstract} 
We report the results of spectroscopic observations, obtained with the 
GEMINI Multi-Object Spectrograph, of 8 planetary nebulae (PNe) in the 
dwarf spheroidal (dSph) galaxy NGC~147, a companion of M~31. The 
physico-chemical properties of the six brightest PNe \citep{Cea05} 
were derived using both the empirical {\sc icf} method and 
photoionization modelling with {\sc cloudy}. Different aspects of the 
evolution of low and intermediate mass stars in a low-metallicity 
environment are analysed using relationships between chemical 
abundances.  In addition, certain features of the chemical evolution 
of NGC~147 were examined. In particular, the mean metallicity of PNe, 
O/H=8.06$_{-0.12}^{+0.09}$ (corresponding to [Fe/H]$_{\rm 
PNe}$$\sim$$-0.97$), is close to the metallicity of the old stellar 
population, [Fe/H]=$-1.0$ \citep[i.e.][]{BMD05}, suggesting a 
negligible chemical enrichment during a substantial amount of time. 
Finally, the luminosity-metallicity relationship for the dwarf 
galaxies of the Local Group is discussed. The location in the 
luminosity-metallicity diagram of dSphs does not exclude their 
formation from old dwarf irregular (dIrs) galaxies, but it does 
exclude their formation from the present time dIrs, since the 
differences between their metallicities are already present in their 
older populations. The offset in the luminosity-metallicity 
relationship indicates a faster enrichment of dSphs, and together with 
the different average abundance ratio [O/Fe] demonstrates the different 
star formation histories for these two types of galaxies. 
\end{abstract} 
 
 
\section[]{Introduction} 
 
The rich variety of galaxies in the Local Group (LG) provides an 
opportunity to study in detail the formation and evolution of the most 
common types of galaxies in the Universe. One remarkable open question 
related to the evolution of dwarf galaxies is: are dwarf ellipticals 
and spheroidals the evolved descendants of previous star-forming 
dwarfs? This hypothesis can be tested through the 
metallicity-luminosity relation for LG dwarfs \citep{Skea89}. The key 
problem with this relation is that there are large uncertainties in 
the [O/Fe] assumed for non star-forming galaxies \citep{RM95}. A way 
to address this problem is by deriving the chemical abundances of a 
significant sample of LG dwarf galaxies using emission-line objects, 
such as PNe, which are present from early- to late-type galaxies. In 
fact PNe and \hii\ regions have been used to address this issue by a 
few groups \citep[]{RM95,Sea98,Mea05}. Our motivation for this paper 
is to contribute to such a study by discussing new spectroscopic 
data for NGC~147 from the GEMINI telescope. 
 
NGC~147, together with NGC~205 and NGC~185, is one of the three  
brightest dwarf companions of M31. At variance with NGC~185, NGC~147 
is gas and dust free. The amount of \hi\ in NGC~205 and NGC~185 is 
significantly higher than in NGC~147, in which the \hi\ column density 
is at least a factor of 10 lower than in the other two galaxies. The 
dust clouds that are prominent in NGC~205 and NGC~185 are also lacking 
in NGC~147 \citep{YL97}. NGC~147 and NGC~185 have similar star 
formation histories and are dominated by their old stellar 
populations, while NGC~205 shows several peculiarities, including a 
younger population \citep{Ma98}. 
 
The stellar metallicity of NGC~147 is relatively well known. Based on 
deep V and I photometry of the central regions, \citet{Da94} has shown 
that NGC~147 is moderately metal-poor ([Fe/H]=$-1$$\pm$$0.3$).  Its 
colour-magnitude diagram indicates the presence of old red giant 
branch (RGB) stars, as well as some stars formed at intermediate 
epochs, like asymptotic giant branch (AGB) stars about 5~Gyr 
old. Using RGB colours, \citet{Hea97} give two slightly different 
metallicities for the inner (up to 1.5$\arcmin$ from the centre) and 
outer (from 1.5 to 4$\arcmin$) fields of the galaxy, namely $-0.91$ 
and $-1.0$, respectively. They point out that there is a weak tendency 
of increasing metallicity with galactocentric radius in the outer 
field.  \citet{Hea97} also stress the presence of an AGB population 
with an age of several Gyrs, and the absence of main sequence stars 
with M$_V$ $<$1, suggesting that the star formation in NGC~147 ceased 
at least 1~Gyr ago. On the other hand, \citet{Nea03} used a 
four-colour photometry to analyse the late-type stellar content of 
this galaxy and that of NGC~185, separating AGB stars of different 
chemistry (O- and C-rich, i.e., M- and C-type stars).  From the mean 
RGB colour indices, they conclude that the metallicity of NGC~147 is 
[Fe/H]=$-1.11$ and that of NGC~185 is [Fe/H]=$-0.89$. 
 
In summary, following the above arguments and the work by 
\citet{Ma98}, while NGC~147 and NGC~185 are very similar in terms of 
their stellar content, star formation history and absolute luminosity, 
their metallicity is different. 
 
In this paper, we present the determination of chemical abundances in 
NCG~147 using PNe. These data allow us to address some aspects of the 
chemical evolution of the galaxy, its consequences for the 
luminosity-metallicity relation of the LG dwarfs, as well as the 
properties of the PNe themselves. In fact, it appears that NGC~147 and 
NGC~185 also have different PNe populations: PNe in NGC~185 are 
systematically brighter than those in NGC~147 \citep{Cea05}.  Note 
that lower limits for the O/H abundances of the PNe in NGC~185 and 
NGC~205 were determined by \citet{RM95}. 
 
The paper is organized as follows. The Multi Object Spectroscopy (MOS) 
data and their reduction are discussed in Section~2. In Section~3, we 
present the derivation and analysis of the PNe physical parameters and 
chemical abundances. These are then discussed in Sect.~4 using 
different relationships, and in Sect.~5 they are compared with the 
stellar metallicity and the age-metallicity relation. Section~6 is 
devoted to the luminosity-metallicity relation for dwarf spheroidals 
and irregulars. Finally, our conclusions are presented in Section~7. 
 
\section[]{Data Acquisition and Reduction} 
 
Seven of the nine PNe identified in NGC~147 from narrow band 
photometry by \citet{Cea05} were observed spectroscopically using the 
GEMINI Multi-Object Spectrograph North (GMOS-N). 
 
\begin{figure*} 
\centering \vbox{ \psfig{file=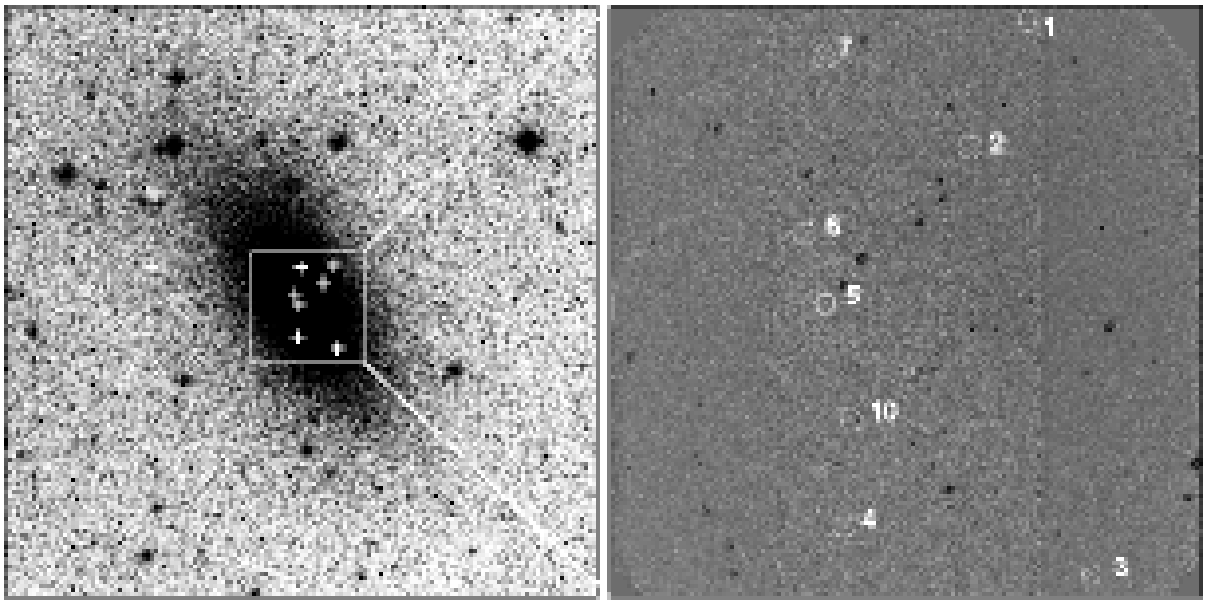,width=16.0truecm}} 
\caption{{\it Left}: The 34$\arcmin \times$34$\arcmin$ Digitalized 
Sky Survey image centred in NGC~147. Plus signs mark the position 
of the PNe candidates studied by \citet{Cea05}. {\it Right}: The 
5.5$\arcmin \times$5.5$\arcmin$ GMOS-N continuum-subtracted, 
\ha-R, image of NGC~147, with circles marking the position of the 
8 PNe contained in the MOS mask. See text and Table~1 of 
\citet{Cea05} for the RA and DEC of the PNe.} \label{GMOS_im} 
\end{figure*}

The pre-imaging needed to build the MOS mask was also obtained with 
GMOS-N on August 14, 2005. Images were taken using a narrow-band \ha\ 
filter and a broad-band continuum filter R, whose central $\lambda$ 
and $\Delta\lambda$ are 655/6.9nm and 657/83nm, respectively. The 
exposure times were 2$\times$300s in \ha\ and 2$\times$60s in R. The 
seeing was $<$0.7~arcsec. These images were used to produce a 
continuum-subtracted image to determine the position of the long-slits 
and to create a MOS mask containing seven PNe. In addition, a new PN 
was identified with the GMOS pre-imaging and consequently inserted in 
the spectroscopic observations. The identification of the original 7 
PNe in the MOS mask are those given by \citet{Cea05} in their Table~1, 
and named PN1 to PN7.  The coordinates of the new PN, called PN 10, 
are: R.A.=00 33 11.87, Dec=+48 28 11.7 (J2000.0). Figure~\ref{GMOS_im} 
shows the image of NGC~147, marking the position of the 8 PNe studied 
in this work. 
 
The GMOS-N in its spectroscopic mode was used in two different nights 
to obtain the PN spectra. The ``red" spectra were taken on October 31, 
2005 with the R400+5305 grating, in two exposures of 1800s each. As 
for the ``blue" spectra, data were obtained on December 3, 2005, with 
the B600+G5303 grating, in 2$\times$2800s exposures. In both 
configurations the slits widths were 1~arcsec, and the pixel binning 
was 2$\times$2 (spectral$\times$spatial). The spectral and spatial 
resolutions were: 0.3~nm and 0$''$.094~pix$^{-1}$, respectively, with 
B600 and 0.8~nm and 0.$''$134~pix$^{-1}$ with R400.  The seeing during 
the ``blue'' and ``red'' spectroscopic observations was $<$0.5~arcsec 
and $<$0.7~arcsec, respectively. 
 
A CuAr calibration was obtained with the B600+G5303 grating, but the 
only CuAr exposure taken with R400+5305 was not good enough to be used 
for wavelength calibration. Because of that, we have calibrated all 
the ``red" spectra using a list of skylines. This method gives a 
satisfactory wavelength calibration, thanks to the large number of 
skylines in the red part of the spectra. An exposure of the 
spectroscopic standard star G191B2B was taken with each configuration 
around the date of the observation of the spectra, and these exposures 
were applied in the flux calibration of our data. Data were reduced 
using the {\sc Gemini GMOS data reduction script} and {\sc longslit} 
tasks, both being part of the IRAF package. 
 
The selected slit length varies from one PN to another according to 
the field crowdedness. Also, different slit distances from the 
bisector of the mask in the North-South direction correspond to a 
different wavelength coverage. Typically, the ``blue" spectra covered 
a spectral range from 300--350~nm to 570--650~nm. In the case of the 
``red" spectra, they covered from 500-620~nm to 920--1050~nm. Note 
that there is some wavelength overlap between the 
``blue" and ``red" spectra. As an intense emission line was not always 
found in the overlap zone, we used the spectra of PN3 which contains 
the \ha\ line in both ``red" and ``blue" spectra to obtain a scale 
factor of 0.8 for all PNe. This scale factor was applied to the fluxes 
measured in the ``red" spectra in order to get the fluxes reported in 
Table~1. In fact, from our previous experience with multi-object spectroscopy, 
we have noticed that differences of the scale factors are generally 
small, amounting to about 5-10\%. We have also evaluated which is the 
error that we are adding to the measured fluxes due to the uncertainty 
in the matching. This error amounts to about 7\%, which can be 
considered a characteristic measure of the uncertainty on the scale 
factors. 
 
Note that in PN5 only the fluxes of the \oiii\ 495.9~nm and 500.7~nm 
lines were measured, so that this PN does not appear in the following 
analysis. Errors in the fluxes were calculated taking into account the 
statistical error in the measurement of the fluxes, as well as 
systematic errors of the flux calibrations, background determination, 
and sky subtraction. Table~2 shows the estimated percentage errors for  
a range of line fluxes (relative to \hb=100) in each PN. Here, and  
throughout the paper, quoted uncertainties are 1-$\sigma$ errors. 
 
\section[]{Data analysis} 
 
\subsection[]{Extinction and Physical Parameters} 
 
The observed line fluxes were corrected for the effect of the 
interstellar extinction using the extinction law by 
\citet{Ma90} with $R_V$=3.1.  We used \cbeta\ as a measurement of the 
extinction, which is defined as the logarithmic difference between 
the observed and de-reddened \hb\ fluxes. 
Since the ``blue" part of the spectra, which contains also the 
\hd\, and \hg\ lines, is affected by larger uncertainties, 
the \cbeta\ was determined comparing only the observed Balmer I(\ha)/I(\hb) 
ratio with its theoretical value, 2.85 \citep{Os89}. 
For the higher signal to noise spectrum of PN4, whose \hb\ line was 
lost in the gap between two CCDs, the I(\ha)/I(\hg) was used to 
compute the extinction. The values of \cbeta\ and their 1-$\sigma$  
errors are shown in the last row of Table~1.  
\cbeta\ is related to E(B-V) through the following relation: 
 \cbeta=0.4 R$_{\beta}$ E(B-V), where R$_{\beta}$=3.7. 
The average value of the extinction of the PNe 
$<$\cbeta$>$=0.07$\pm$0.07 gives E(B-V)=0.05$\pm$0.10, which is 
comparable with the reddening value E(B-V)=0.173 derived by  
\citet{Slea98} toward NGC~147, from dust infrared emission 
features. Patchiness of dust and gas toward NGC~147 \citep{Sea06} 
might be the reason of appreciable variations  of the reddening 
throughout the galaxy area. 
 
The extinction-corrected intensities (given in the middle column of 
Table~1) were then used in order to obtain the densities and 
temperatures for each PN.  As we can see in Table~1,  
\sii$\lambda\lambda$671.6,673.1, usually 
used as an electron density estimator, was measured only in PN4, PN6 
and PN10 (and an upper limit is given for PN1). As for the temperature 
estimation, we were able to use the 
I($\lambda$495.9+$\lambda$500.7)/I($\lambda$436.3) ratio, giving 
\te\oiii, for all the PNe but PN10 (also note that only an upper limit 
is given for PN6). In the particular case of PN4, two other 
temperature estimators were also used, namely  
I($\lambda$654.8+$\lambda$658.3)/I($\lambda$575.5) for \te\nii\ and 
I($\lambda$372.6+$\lambda$372.9)/I($\lambda$732.0+733.0) for 
\te\oii. The derived electron densities and temperatures, 
and their errors, are listed in the upper part of 
Table~3.  
 
This is the first direct determination of the physical parameters of 
emission line objects in NGC~147, which prevents us to perform any 
external comparison.  
 
\subsection[]{Empirical {\sc icf} abundances} 
 
Chemical abundances were derived from the emission-line intensities 
using the Ionization Correction Factors ({\sc icf}) method. The {\sc 
icf} were computed following the prescriptions by \citet{KB94}. 
 
As mentioned above, the \oiii\ 436.3~nm emission line was measured 
with a sufficiently high signal to noise ratio in five PNe, namely 
PN1, 2, 3, 4, and 7. There, \te\oiii\ could be directly determined.  In 
one case, PN6, we could estimate an upper limit to the \oiii\ 436.3~nm 
emission line, and consequently an upper limit to \te\oiii. 
 
For PN~10, we could not measure this temperature diagnostic emission 
line, and therefore we could not derive its chemical abundances. 
Finally, as noted before, in the spectrum of PN5, only the \oiii\ 
doublet 495.9,500.7~nm could be detected, preventing any further 
analysis. In PN4, \te\oiii\ was used in computing chemical abundances 
with the {\sc icf} method for ions which are ionized twice or more 
times, and \te\nii\ for singly ionized species. 
 
The {\sc icf} method was applied to the six PNe with an electron 
temperature measurement. Formal errors on the {\sc icf} abundances 
were computed taking into account the uncertainties in the observed 
fluxes, in the electron temperatures and density, and in the \cbeta, 
as done in \citet{Gea03} and \citet{Mea05}. The ionic and total 
abundances, as well as the ionization correction factors, are reported 
in Table~4.  
 
\subsection[]{{\sc cloudy} photoionization modelling} 
 
PNe were also modelled using the photoionization code {\sc cloudy} 
95.06 \citep{Fea98}.  The complete procedure is described in 
\citet{Mea04}.  Here we remind the most important parameters and 
assumptions. The input parameters for {\sc cloudy} are {\it i}) the 
energy distribution, {\it ii}) the luminosity of the central star, 
{\it iii}) the nebular geometry, {\it iv}) the hydrogen density, 
N$_H$, and {\it v}) the chemical abundances.  At the large distance of 
NGC~147, PNe are not resolved and we assumed the most simple geometry 
of spherical symmetry with constant density.  The energy distribution 
of the central star was set to that of a blackbody (BB). In fact, as 
described by \citet{Mea04}, the model atmospheres by \citet{Ra03} give 
similar results in terms of chemical abundance determinations. Dust 
grains were added to the models, since they have an important effect, 
particularly on the temperature structure of the PNe \citep[see, for 
  instance,][]{Gea04}. This effect depends on the type of grain 
(graphite and/or silicate), the grain abundances, and the grain-size 
distribution, being more relevant in the inner regions of the nebula 
\citep[see][]{DS00}.  Dust type and size distribution adopted are 
those typical of Galactic PNe, with ISM gas-phase depletion due to 
grains following \citet[]{VHea04}. 
 
The BB temperature T$_{\star}$ was derived using Ambartsumian's method 
\citep{A32}, i.e. using the \heii($\lambda$468.6)/{H{\sc 
i}}($\lambda$486.1) line flux ratio, or the ([O{\sc 
iii}]$\lambda$495.9+[O{\sc iii}]$\lambda$500.7)/\heii$\lambda$468.6 
line ratio method \citep{Gu88}. This is done only for the first 
iteration, in the following ones the BB temperature was adjusted as to 
match the \hei/\heii\ line ratio. 
 
The luminosity L$_{\star}$ of the central star was set to reproduce the 
photometric absolute flux of the \oiii\ $\lambda$500.7 nebular line, 
as given in \citet{Cea05}.  The external and internal radii of the 
shell, R$_{in}$ and R$_{ext}$, were varied to match the ratio of lower 
and higher excitation ions, as \oiii/\oii.  Density was derived from 
the \sii\ $\lambda$671.7/$\lambda$673.1 intensity ratio, which was 
measured in PN1, PN4, and PN6. For the remaining PNe, it was assumed 
to be equal to 5000~cm$^{-3}$, which can be considered a typical value 
for Galactic PNe \citep[see e.g. ][]{SK89}.  In addition, we made some 
tests varying the density by $\pm$1000~cm$^{-3}$ with both the {\sc 
icf} and {\sc cloudy} methods, obtaining negligible changes in the 
computed abundances. We assumed the average non-Type~I Galactic PNe 
abundances from \citet{KB94} as initial chemical abundances.  The 
parameters of the best {\sc cloudy} model (for the central star 
T$_{\star}$, L$_{\star}$, and for the nebula N$_H$, R$_{in}$, 
R$_{ext}$) are shown in the second half of Table~3.  
 
Subsequently, chemical abundances were varied to match the observed 
and predicted intensities within 5\% for the lines brighter than \hb\ 
and 10\% for the other lines. As explained above, we considered dust 
in our modeling. This inclusion came from the necessity of finding a 
good agreement between the empirical and {\sc cloudy} 
temperatures. The abundances of C, O and Fe (the most important 
coolants) were also slightly modified to refine the match with the 
empirical T$_{e}$. It is clear from Table~3 that 
the model T$_{e}$ is similar to the empirical one, with the 
exception of PN4. In the latter case, we have three different 
empirical measurements of the electron temperature. Our simple model 
with uniform density is not capable of reproducing the empirical 
temperatures, likely because of different conditions in different 
zones of the nebula. We therefore decided to reproduce just the the 
average of \te\oiii\ and \te\nii. No other simple assumption for the 
\ne\ radial profile (e.g. decreasing with radius) was able to 
reproduce satisfactorily the observed line intensities. 
 
The results of the best-fitting {\sc cloudy} models, in terms of 
chemical abundances (and their errors) are shown in 
Table~5. Note that {\sc icf} abundances are in good agreement with those derived 
using {\sc cloudy}, whose mean values for the different elements, in 
units of 12~+~log~X/H, are: He/H=11.02$^{+0.20}_{-0.36}$; 
O/H=8.06$_{-0.12}^{+0.09}$; N/H=7.70$_{-0.31}^{+0.18}$; 
Ne/H=6.68$_{-0.32}^{+0.18}$; S/H=6.35$_{-0.68}^{+0.25}$; and 
Ar/H=5.13$_{-0.64}^{+0.24}$.  
 
\section[]{Chemical abundance patterns} 
 
PNe are formed by stars ranging from 0.8 to 8 M$_{\odot}$, which loose 
their chemically enriched envelopes during the AGB evolution.   
Relationships between chemical abundances of He/H, N/O, N/H and O/H 
are known to exist in Galactic PNe \citep[cf.][]{He90}.  Extragalactic 
PNe are important because they allow us to test if these relationships 
also hold in different environments.  In particular, PNe belonging to 
dwarf galaxies permit to extend these relations to very low 
metallicities. 
 
\begin{figure} 
\centering 
\vbox{ 
\psfig{file=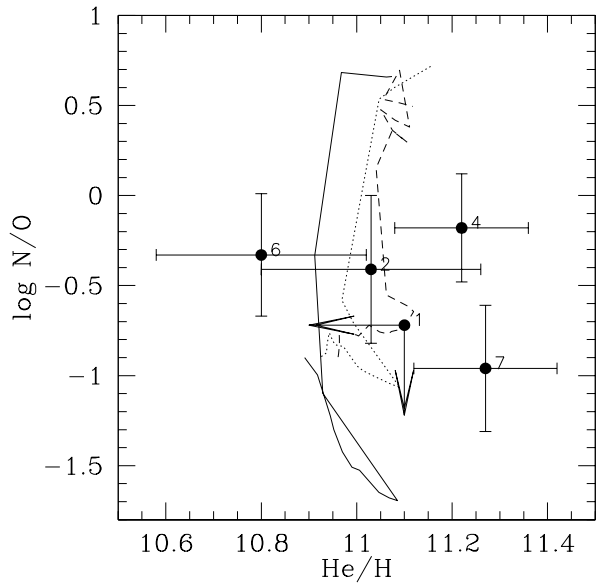,width=8.0truecm}} 
\caption{The relationship between log~N/O and He/H. The solid (Z=0.001) and 
dotted (Z=0.004) lines represent the stellar models for the progenitors 
(Marigo 2004, private communication), and the dashed line is the model with Z=0.008 \citep{marigo01}. 
Filled circles correspond to NGC~147 PNe with {\sc cloudy} He/H and N/O abundances.} 
\label{fig_he_no} 
\end{figure}

The relation between log~N/O and He/H for the PNe of NGC~147 (except 
PN3, for which only a crude limit of N/H could be derived) is shown in 
Figure~\ref{fig_he_no}, where our {\sc cloudy} chemical abundances are 
compared with the models by \citet{marigo01}, built for initial 
metallicities of Z=0.008, and new models by Marigo (private 
communication) for Z=0.001 and Z=0.004.  The lower metallicity models 
(Z=0.001 and Z=0.004) predict, for progenitor masses below 
3M$_{\odot}$, an enhancement of O/H compared to N/H, producing a N/O 
abundance ratio that decreases with He/H. For  higher mass 
progenitors, N/O increases quickly and significantly as the result of 
hot-bottom burning.  In the model with Z=0.008, the N/O ratio remains 
almost constant for initial masses below 3M$_{\odot}$.  The metallicity 
of NGC~147 is Z=0.002 \citep[cf.  ][]{Da94,Hea97,Nea03}.  Taking into 
account the errors of the He abundances and of the N/O ratio, the 
NGC~147 PNe are consistent with the Z=0.001 and Z=0.004 tracks.  PN4 
and PN7 might be consistent with a higher metallicity, and 
consequently a more recent formation. 
 
\begin{figure} 
\centering \vbox{ \psfig{file=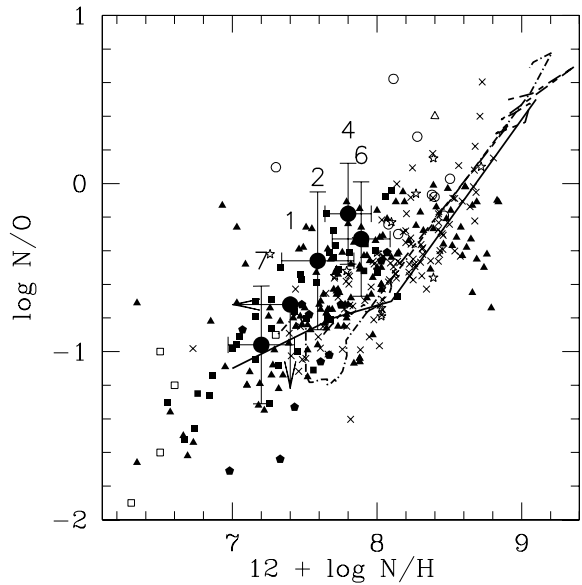,width=8.0truecm}} 
\caption{The relationship between log~N/O and 12~+~log~N/H for LG 
PNe.  The dash-dotted line is the model by \citet{groene} for the 
LMC metallicity and initial masses from 0.93 (left) to 8.5 
M$_\odot$(right).  The solid line indicates the model for Z=0.004 
and the dashed line that for Z=0.008. Both models are for initial 
masses from 0.8 (left) to 5~M$_\odot$(right) \citep{marigo01}. The 
{\sc cloudy} abundances of NGC147 PNe are marked with filled 
circles. Observed abundances are from: Galaxy \citep[crosses, 
][]{peri04}; LMC and SMC  \citep[filled triangle and squares, 
respectively, ][]{leisy06}; M31 \citep[stars, ][]{jc99}; M33 
\citep[filled pentagons, ][]{Mea04}; Sextans A and B \citep[empty 
triangle and squares, ][]{Mea05}; M32 \citep[empty circles, 
][]{Rea99}.} \label{fig_nh_no} 
\end{figure}

The correlation between log~N/O and 12~+~log~N/H is shown in 
Figure~\ref{fig_nh_no} for PNe belonging to several LG galaxies.  This 
relation is well defined for PNe with a slope close to unity, 
suggesting that the increase of N/O with N/H is due to the increase of 
N, which is mainly produced by the CN cycle, without modifying the 
oxygen abundance through the ON cycle.  Theoretical predictions of N/O 
vs. N/H for different masses of the progenitor stars and metallicity 
(Groenewegen \& de Jong 1994, for the LMC metallicity; and Marigo 
2001; for Z=0.001 and Z=0.004) are also shown.  Note that Type~I PNe 
which experienced hot bottom burning conversion of C to N, are located 
outside the relation.  The PNe of NGC~147 are in agreement with the 
general behaviour of the LG PNe.  Following the classical definitions 
of Galactic Type~I PNe, N/O$\ge$$-0.3$ and He/H$\ge$$11.1$ 
\citep{peimbert} or the more stringent one N/O$>$$-0.1$ \citep{KB94}, 
the NGC~147 sample does not include Type~I PNe.  However, it has to be 
noted that the original definition of Type~I PNe only holds for the 
metallicity of the Galaxy \citep{Mea04}. The absence of Type~I PNe 
would not be surprising if we consider that they are typically 
associated to younger progenitors \citep{CS95} while the predominant 
stellar population in this galaxy is very old. 
 
\begin{figure} 
\centering \vbox{ \psfig{file=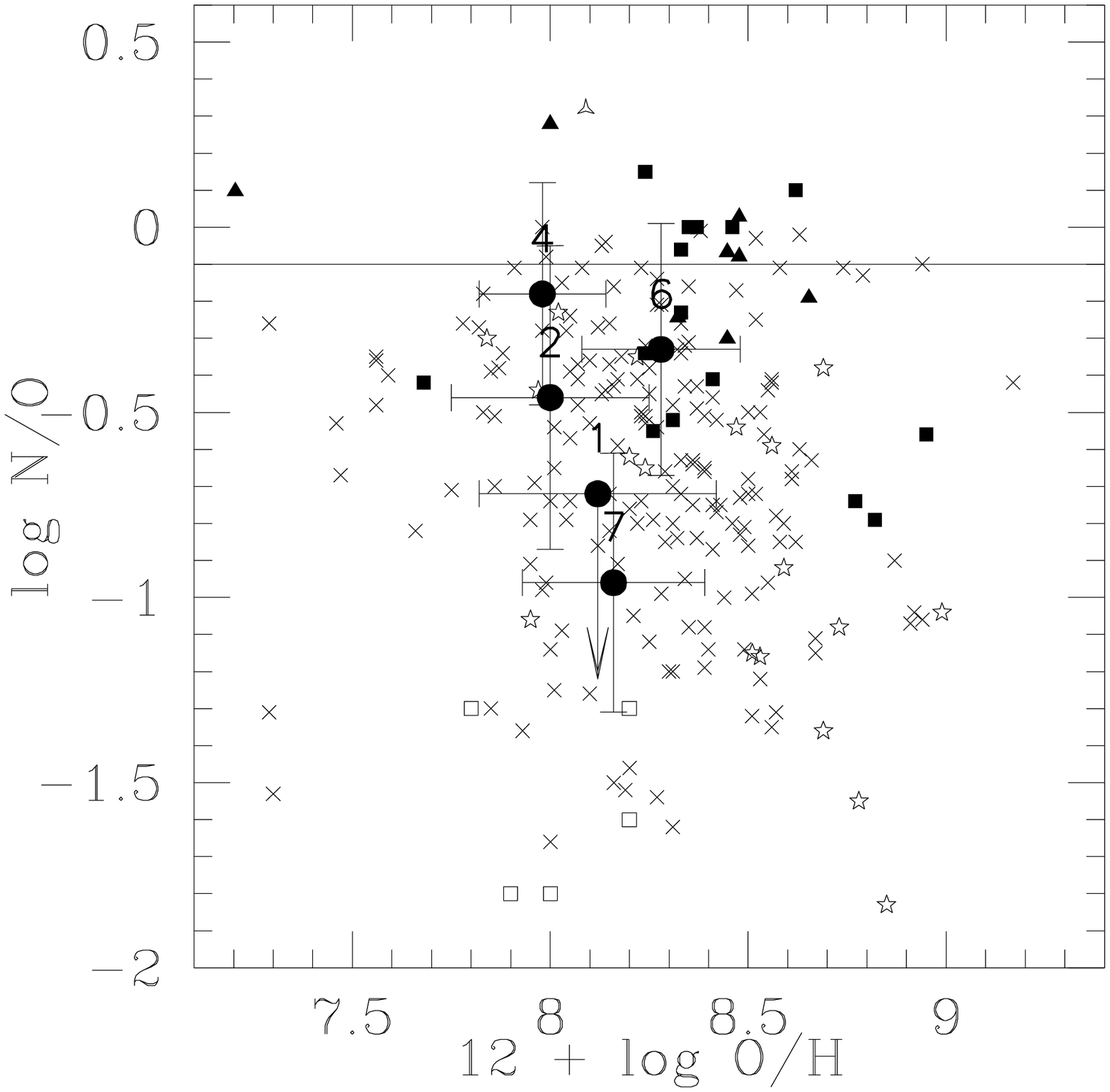,width=8.0truecm}} 
\caption{log~O/H vs. 12~+~log~N/O. {\sc cloudy} abundances of the 
PNe in NGC~147 are marked with circles.  Observed abundances are 
from: M31 bulge \citep[filled squares, ][]{jc99}; M33 \citep[empty 
stars, ][]{Mea04}; Magellanic Clouds \citep[crosses, ][]{leisy06}; 
M32 \citep[filled triangles, ][]{Rea99}; Sextans A and B 
\citep[empty triangle and squares, ][]{Mea05}; LeoA \citep[cross, 
][]{Vea06}.} \label{fig_ohno} 
\end{figure}

Figure~\ref{fig_ohno} shows the log~O/H vs. 12~+~log~N/O for a sample of LG PNe. 
There is not a clear correlation between these two quantities, as 
already noted in previous works \citep[cf. ][]{He90,R06}. 
 
\section[]{The chemical evolution of NGC~147} 
 
The elemental abundances shown in Table~4 and Table~5 can also be compared  
with the metallicity measured in stars of this galaxy.  Given that {\sc icf} 
and {\sc cloudy} abundances are in good agreement, we use the {\sc 
cloudy} O/H mean value, as given in the last row of Table~5,  
that considers six PNe. 
 
In principle, O/H can be converted into [Fe/H] using the 
following relation: [Fe/H]$_{\rm PNe}$=[O/H]$_{\rm PNe}$$- 0.37$ 
obtained by \citet{Ma98}. This relation is useful for comparing 
samples of galaxies, but less valid considering galaxies individually. 
Even on a statistical sense, this transformation is uncertain 
\citep[$\pm0.06$~dex;][]{Ma98}. Keeping in mind this strong 
limitation, the mean O/H that we obtained for NGC~147, gives 
[Fe/H]$_{\rm PNe}$=$-0.97$ \citep[[O/H$_{\rm solar}$=8.66;][]{As03}, 
in agreement with metallicities of the RGB population 
\citep{Da94,Hea97,Nea03}, discussed in Sect.~1. Moreover, when we plot 
[Fe/H]$_{\rm PNe}$ in terms of the galactocentric distance (not shown 
here), the weak gradient with metallicities increasing from the 
central to the outer zones of the galaxy, argued by \citet{Hea97}, is 
not confirmed. In summary, this crude agreement between PNe 
metallicities and old-population metallicity would suggest that 
NGC~147 did not experience a significant chemical enrichment during a 
long period of time. 
 
\begin{figure} 
\centering 
\vbox{ 
\psfig{file=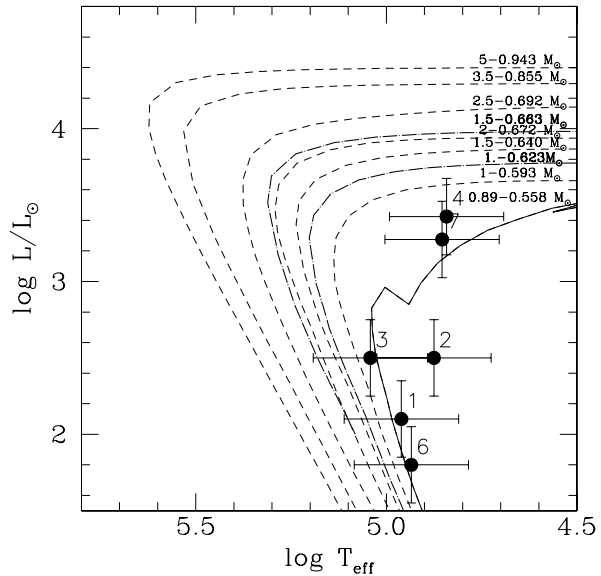,width=8.0truecm}} 
\caption{The H-R diagram of the PNe central stars. Luminosities and temperatures were 
derived with the {\sc cloudy} photoionization models. The evolutionary 
tracks are by \citet{VW94}: Z=0.004 continuous lines and Z=0.001 dashed lines.} 
\label{fig_track} 
\end{figure}

As described in Section 3.3, the modelling with {\sc cloudy} allowed 
us to estimate stellar parameters as the luminosity and temperature of 
the central stars. The model T$_{\star}$ and L$_{\star}$ are listed in 
Table~4 and can be located in the $\log$T$_{\star}$-$\log$L$_{\star}$ 
diagram. Their location are then compared with the available H-burning 
and He-burning evolutionary tracks, for Z=$0.004$ and Z=$0.001$ 
\citep{VW94}, as shown in Figure~\ref{fig_track}. From the loci of the 
central stars in the H-R diagram, we can estimate the central star 
masses, which range from 0.56 to 0.59~M$_{\odot}$. Using the initial 
to final mass relationship adopted by \citet{VW94}, we find that the 
masses of the progenitor stars were between 0.9 and 
1.0~M$_{\odot}$. Stars with these main sequence masses and with an 
initial metallicity of Z=0.001 were born about 10 to 7~Gyr ago 
\citep{charbo99}. This is in agreement with what is known about the 
star formation history of NGC~147.  As described in Sect.~1, the 
colour-magnitude diagrams suggest the presence of old RGB stars and of 
a small number of intermediate-age AGB stars with an age of about 5~Gyr. 
 
From the study of chemical evolution of dwarf spheroidal galaxies, it 
is known that these galaxies are characterized by one or two long 
bursts of star formation \citep{LM03}.  The low gas content is the 
result of gas consumption by star formation and gas removal by an 
efficient galactic wind. 
 
A detailed comparison of the PNe abundances obtained in this paper 
with those for the other two bright dwarf companions of M31 is not 
possible. The only other available PNe study for these galaxies is in 
fact that of \citet{RM95}, who have observed only two PNe in NGC~205, 
one of them giving 12~+~log~O/H=8.21 and the other $\ge$ 8.11. In 
the case of NGC~185 \citep{RM95}, the \oiii4363~nm line was not 
detected in any of the PNe, thus the derived abundances are quite 
uncertain. \citet{RM95} applied empirical methods, derived from 
properties of PNe in the Magellanic Clouds and the Milk Way, to 
correct their lower limits of O/H, concluding that their values 
underestimate the O/H abundances of these two galaxies by 
0.37~dex. After this correction, they obtained 12~log~O/H=8.60 and 
8.20 for NGC~205 and NGC~185, respectively, which are comparable to 
the value of 8.06$_{-0.12}^{+0.09}$ that we find for NGC~147 (within 
the uncertainties). 
 
\section[]{The luminosity-metallicity relationship of \di\ and \ds} 
 
It has been proposed that dwarf spheroidal galaxies are formed through the 
removal of the gas in dwarf irregulars, either through ram pressure 
stripping, supernova driven winds or star formation \citep{KD89}.  
In this scenario, as described by \citet{RM95}, once the gas finished  
and the star formation stopped, the old dIr-like galaxies  
fade in luminosity as their stellar population ages.  According to this, dSphs  
are expected to have, at a given luminosity, a higher metallicity than dIrs.  
The luminosity-metallicity relationship has been deeply studied for dIr 
galaxies by several authors \citep[]{Skea89, RM95, Lea01, Vea06, 
Lea06}. One of the most recent compilation of oxygen abundance in 
nearby low luminosity gas rich galaxies, dIrs, is by \citet{Vea06}. 
The weighted least-squares fit to the luminosity vs. metallicity 
data for the fifty dIr galaxies within 50 Mpc with an oxygen 
abundance determination is 
\begin{equation} 
12+\log(O/H)=5.67  - 0.151 M_B. 
\end{equation} 
While oxygen abundances can be derived relatively easily for gas rich 
galaxies using \hii\ regions, it is quite difficult to obtain them for 
non-star forming dSph galaxies.  On the other hand, for these galaxies 
the stellar iron abundances are known.  The comparison of dSph 
metallicity with dIr oxygen abundances is thus dependent on the [O/Fe] 
adopted. This ratio depends in turn on the star formation history 
\citep{GW91}.  To overcome the problem, \citet{RM95} proposed to use 
PNe as probes of the stellar oxygen abundance in both dIrs and dSphs. 
PNe are in fact present in all types of galaxies and, with few 
exceptions \citet[][cf. Sextans A PNe]{Mea05}, they do not produce 
oxygen, and thus they are representative of the stellar progenitor 
oxygen abundance. 
 
\begin{figure} 
\centering \vbox{ \psfig{file=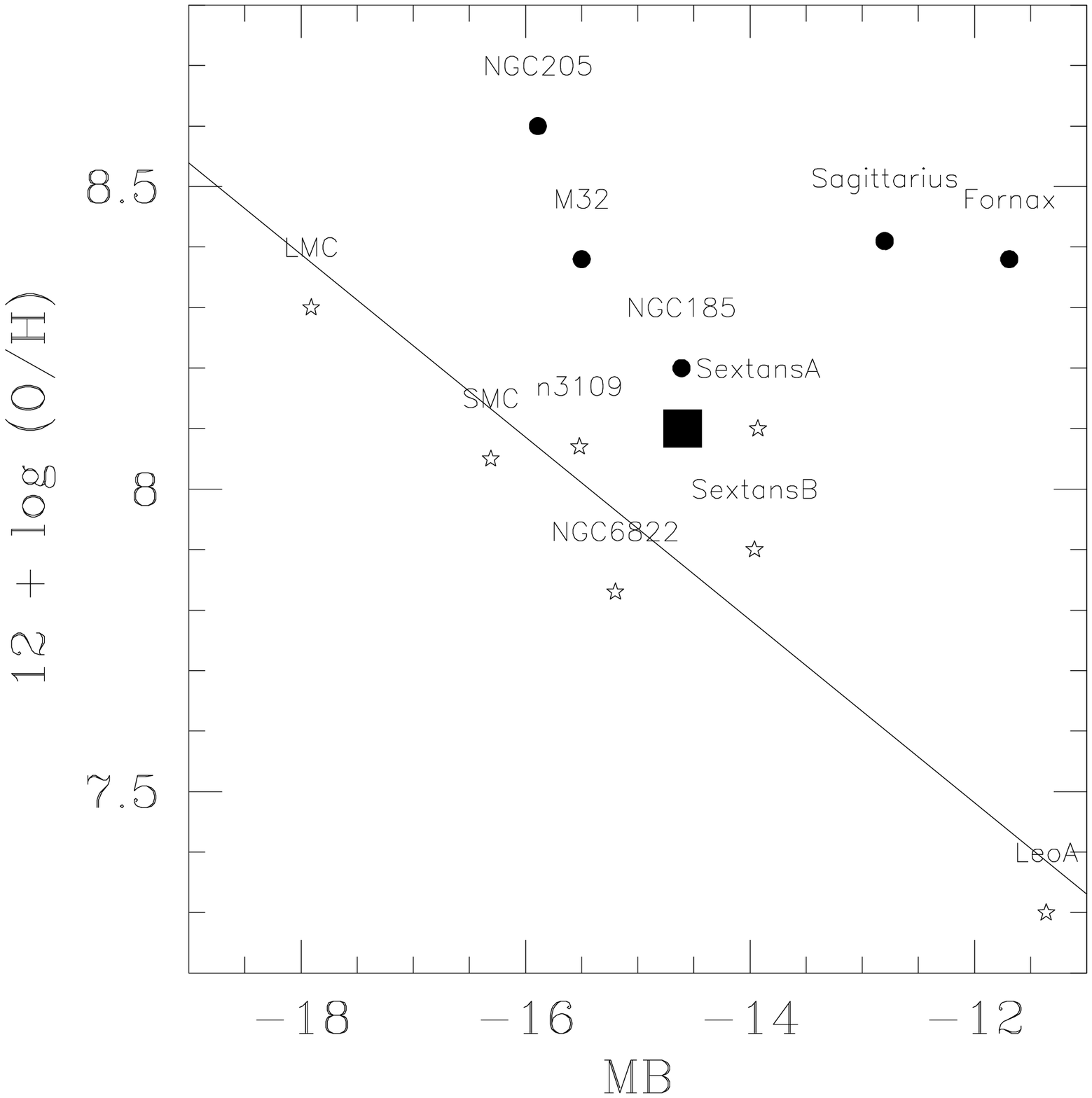,width=8.0truecm}} 
\caption{Luminosity-metallicity for dwarf galaxies in the LG. 
dSphs are marked with filled circles, dIrs with stars, NGC~147 
average abundance with a filled square. Abundances determinations: 
NGC~147 (this work), LMC and SMC \citep{leisy06}, Sextans~A and B 
\citep{Mea05}, NGC~205, NGC~185, Fornax \citep{RM95}, Sagittarius 
\citep{zijlstra06}, Leo~A \citep{Vea06}, M32 \citep{Sea98}, 
NGC~6822, NGC~3109 \citep{leisy07}. The almost diagonal line marks 
the luminosity-metallicity relation obtained from \hii\ regions 
\citep{Vea06}.} \label{fig_lummet} 
\end{figure}

Figure~\ref{fig_lummet} shows the luminosity-metallicity relationship 
obtained using oxygen abundances of \hii\ regions. the continuous line 
is the fit by \citet{Vea06}.  The oxygen abundances of PNe in dIrs are 
marked with stars and those of PNe in dSphs with filled circles. The 
average oxygen abundance of PNe in NGC~147 is marked with a filled 
square.  Note that the dIr PNe abundances are in good agreement with 
the luminosity-metallicity relation from \hii\ regions, with the 
exception of the only PN of Sextans A, where oxygen dredge-up occurred 
\citep[see][]{Mea05}. On the other hand, the oxygen abundances of 
dSph PNe exceed those predicted by eq.~1: at a given luminosity, dSphs 
have a higher metallicity than dIrs. 
 
The similarity of the oxygen abundance found in PNe belonging to 
dIr galaxies to that found in the ISM (\hii\ regions) suggests that 
the observed PNe in dIrs, which are also the brightest, are the 
products of rather recent star formation.  At variance, the mean 
abundance in the PN population of dSph galaxies, where star formation 
ceased long ago, should be closer to the mean abundance of the stellar 
populations in that galaxy. Considering, thus, the difference of the 
age of the PN populations in dIrs and dSphs, the true difference 
between the oxygen abundances referred to the same epoch in dwarf 
irregulars and spheroidals might be larger than what Figure~6 
indicates.  The offset in the luminosity-metallicity diagram does not 
rule out the possibility that dSph galaxies come from primordial 
dIrs-like progenitors that consumed all their gas content and then 
faded in luminosity \citep[cf. ][]{RM95}. On the other hand, the 
metallicity offset is against the hypothesis that dwarf spheroidals 
form from present-time dwarf irregulars. 
As discussed by \citet{grebel05}, dwarf irregulars and dwarf 
spheroidals may have been very similar initially, but the early 
evolution of the two types differed greatly, with dwarf spheroidals 
forming more stars and enriching themselves more, as observed from 
their stellar population --which is more metal rich than those of 
dIrs. 
 
If we also consider the [O/Fe] ratio, as discussed extensively by 
\citet{RM95}, large differences of this ratio in dIrs and dSphs are 
found (see their Table 6).  [O/Fe] is larger in dSphs than in dIrs, 
reflecting the higher gas consumption rate of the \ds\ and therefore a 
substantially different star formation history.  The [O/Fe]=0.44 of 
NGC~147, computed adopting [Fe/H]=-1 by \citet{Da94} and solar 
abundances of iron and oxygen by \citet{As03}, point in this 
direction. 
 
In conclusion,  
the offset in the luminosity-metallicity relationship indicates a 
faster enrichment of dSphs, and together with the different average 
abundance ratios [O/Fe] highlights different star formation histories 
for these two classes of galaxies. 
 
\section[]{Conclusions} 
 
Using the {\sc icf} method as well as {\sc cloudy} photoionization 
modelling, we have analyzed new GEMINI MOS data for the PNe of 
NGC~147, one of the most remarkable dwarf spheroidal companions of 
Andromeda. The physical parameters and the elemental abundances of He, 
O, N, Ne, Ar and S, for the six brightest PNe identified in this 
galaxy by \citet{Cea05} have been determined.  
 
We find that, adopting the definition in the Galaxy, there are no 
Type~I PNe in the NGC~147, an expected result, since Type~I PNe are 
typically associated to younger and massive progenitors, while the 
predominant stellar population in this galaxy is old. 
 
From the literature, the comparison of the near-IR photometry of the 
AGB stars in the three brightest M31 dwarf companions shows that ages 
of 1 and 0.1~Gyr are predicted for the most recent star formation in 
NGC~185 and NGC~205, respectively, while the last event that formed 
stars in NGC~147 occurred some 1 to 3~Gyr ago \citep{Da05,Hea97}. In 
terms of the stellar metallicity, the three galaxies have similar 
median photometric abundances: [Fe/H]=-1.11$\pm$0.08 (NGC~185); 
-1.06$\pm$0.04 (NGC~205); and -1$\pm$0.03 (NGC~147) \citep{BMD05}. 
 
Our determination of the oxygen abundance for the PNe of NGC~147, is 
comparable (within the uncertainties) to those of NGC~185 and 
NGC~205. The O/H abundance converts to [Fe/H]$_{\rm PNe}$=-0.97, which 
is similar to the stellar metallicity of NGC~147. This similarity 
between intermediate-age PNe population and the old stellar population 
of the galaxy implies that no significant chemical enrichment has 
occurred during a long period of time. 
 
Finally, the luminosity-metallicity for the dwarf galaxies of the 
Local Group is analyzed. We conclude that the location in the 
luminosity-metallicity diagram of dSphs exclude their formation from 
present-time dIrs, and together with the abundance ratio [O/Fe] points 
to different star formation histories for these two classes of 
galaxies. 
 
\section*{Acknowledgments} 
 
We would like to thank M. Richer, the referee of this paper, for his 
valuable comments and suggestions which significantly improved the 
paper.  We acknowledge M. Perinotto for using his program to perform a 
fast evaluation of the chemical abundances that we then compared with 
our own evaluation.  The work of DRG is supported by the Brazilian 
Agency FAPESP (03/09692-0 and 04/11837-0), and that of LM  
by the Italian National Institute of Astrophysiscs (INAF).

\begin{table*} 
\begin{center} 
{\scriptsize 
\caption{Observed emission-line fluxes, de-reddened intensities and {\sc cloudy} 
modeled intensities} 
\begin{tabular}{l| lll| lll| lll| lll} 
\hline\noalign{\smallskip} 
 & \multicolumn{3}{c}{PN1} & \multicolumn{3}{c}{PN2} & \multicolumn{3}{c}{PN3} & \multicolumn{3}{c}{PN4}\\ 
\\ 
Ion           & F      &I      &I$_{\rm CLOUDY}$ &  F    &I      &I$_{\rm CLOUDY}$ &  F    &I      &I$_{\rm CLOUDY}$ &  F    &I            &I$_{\rm CLOUDY}$   \\ 
\hline 
\noalign{\smallskip} 
372.7   \oii\     & $<$5.3 &$<$5.3 & 15.0         &$<$8.3 &$<$8.5 & 30.        &-      &-      &-           &27.5        &30.1            &30.1\\ 
383.5    H9       & -      &-      & -            &-      &-      & -           &-      &-      &-       &4.5     &4.9         &7.3\\ 
386.8   \neiii\   & -      &-      & -            &30.4   & 30.6  & 30.2        &8.6    &8.6    &8.4         &9.7     &10.5        &10.5\\ 
388.9    \hei, H8 & -      &-      & -            &-      &-      & -           &73.6   &73.6   &-       &8.5     &9.2         &10.5\\ 
396.7    \neiii,H7& -      &-      & -            &58.2   &59.5   & -           &62.9   &62.9   &-       &17.8    &19.2        &16.0\\ 
410.1    \hd\     &  -     &-      & -            &-      &-      & -           &-      &-      &-       &23.6    &25.1        &26.0\\ 
434.0    \hg\     & 36.9   &37.1   & 47.1         &41.5   &42.0   & 47.2        &33.8   &33.8   &46.6        &44.1    &46.0        &47.0\\ 
436.3    \oiii\   & 11.5   &11.6   & 9.5         &17.4   &17.6   & 16.0        &16.0   &16.0   &15.5         &1.5     &1.6         &3.7\\ 
447.1    \hei     & -      &-      & -            &-      &-      &             &-      &-      &-       &4.0     &4.1         &7.9\\ 
468.6    \heii    & $<$10  &$<$11  & 16.6         &3.5    &3.5    & 4.9         &14.1   &14.1   &13.4        &1.7     &1.7         &1.8\\ 
486.1    \hb\     & 100.   &100.   & 100.         &100.   &100    & 100         &100.   &100    &100         &100.    &100.        &100.\\ 
495.9    \oiii\   & 244.4  &244.1  & 235.1        &308.3  &307.6  & 291.0       &255.4  &255.4  &242.1       &120.4   &119.5       &121.4\\ 
500.7   \oiii\    & 710.7  &709.6  & 707.7        &865.6  &862.6  & 875.4       &712.2  &712.2  &729.0       &364.    &359.8       &365.3\\ 
575.5   \nii\     & -      &-      & -            &-      &-      & -           &-      &-      &-       &1.6     &1.5         &0.8\\ 
587.6   \hei\     & 14.5   &14.4   & 14.0         &13.0   & 12.7  & 12.4        &6.4    &6.4    &6.7         &22.9    &21.4        &21.2\\ 
654.8   \nii\     & -      &-      & -            &-      &-      & -           &-      &-      &-       &17.4    &15.8        &15.1\\ 
656.3   \ha\      & 288.9  &285.0  & 278.6        &293.6  & 285.0 & 274.0       &260.   &260.   &258.        &315.    &285.        &281.\\ 
658.4   \nii\     & $<$6   &$<$6   & 6.0          &17.1   & 16.6  & 16.2        &$<$1   &$<$1   &1.1         &49.6    &44.8        &44.5\\ 
667.8   \hei\     & -      &-      & -            &-      &-      &             &-      &-      &-       &2.5     &2.2         &6.0\\ 
671.7   \sii\     & $<$1.3 &$<$1.2 & 1.2          &-      &-      & -           &-      &-      &-       &1.4     &1.3         &1.3\\ 
673.1   \sii\     & $<$2   &$<$2   & 2.0          &-      &-      & -           &-      &-      &-       &2.1     &1.9         &2.0\\ 
706.5   \hei\     &  -      &-     & -            &-      &-      & -           &-      &-      &-       &1.8     &1.6         &1.6\\ 
713.5   \ariii\   &  -      &-     & -            & 1.3   &1.2    & 1.3         &1.     &1.     & 1.1        &1.6     &1.4         &1.5\\ 
728.1   \hei\     &  -      &-     & -            &  -    &-      & -           &-      &-      &-       &0.5     &0.5         &0.6\\ 
732.0    \oii\    &  -      &-     & -            &  -    &-      & -           &-      &-      &-       &2.4     &2.1         &1.2*\\ 
733.0    \oii\    &  -      &-     & -            &  -    &-      & -           &-      &-      &-       &1.4     &1.2         &1.2*\\ 
906.9    \siii\   & -       &-     & -            &0.57   & 0.54  & 0.61        &-      &-      &-       &-       &-           &-\\ 
953.2   \siii\    & -       &-     & -            & -     &-      & -           &$<$2   &$<$2   & 0.3        &-           &-           &-\\ 
\noalign{\smallskip} 
\cbeta            &0.02$\pm$0.003   &       &     & 0.04$\pm$0.01   &      &     &0.0$\pm$0.02   &       &    &0.14$\pm$0.02         &         &\\ 
\hline 
\noalign{\smallskip} 
\end{tabular} 
} 
\end{center} 
\label{FluxInt} 
\end{table*}

\begin{table*} 
\begin{center} 
{\scriptsize 
\begin{tabular}{l| lll| lll| lll|} 
\noalign{\smallskip} 
\hline\noalign{\smallskip} 
 & \multicolumn{3}{c}{PN6} & \multicolumn{3}{c}{PN7} & \multicolumn{3}{c}{PN10} \\ 
\\ 
Ion            & F      &I      &I$_{\rm CLOUDY}$ &  F    &I      &I$_{\rm CLOUDY}$ &  F    &I      &I$_{\rm CLOUDY}$  \\ 
\hline 
\noalign{\smallskip} 
372.7   \oii\      &$<$35   &$<$35  &74.0         &$<$5   &$<$5   & 17.0         &$<$65  &$<$73  & -          \\ 
396.7    \neiii,H7 & -      &-      &-            &21.6   &22.7   & 16.0        &-      &-      &-       \\ 
410.1    \hd\      &  -     &-      &-            &14.2   &14.8   & 26.1        &-      &-      &-       \\ 
434.0    \hg\      & -      &-      &-            &34.3   &35.3   & 47.0        &-      &-      &-       \\ 
436.3    \oiii\    &$<$12   &$<$12  &7.0          &6.8    &7.0    & 7.7        &-      &-      &-        \\ 
447.1    \hei      & -      &-      &-            &-      &-      & -           &-      &-      &-       \\ 
468.6    \heii     & 5.8    &5.8    &6.3          &4.0    &4.1    & 4.8        &-      &-      &-        \\ 
486.1    \hb\      & 100.   &100.   &100.         &100.   &100.   &100.         &100.   &100    & -          \\ 
495.9    \oiii\    & 186.9  &186.9  &216.2        &257.8  &256.5  & 253.        &200.4  &198.7  & -      \\ 
500.7   \oiii\     & 659.5  &659.5  &650.1        &788.3  &782.0  & 761.5       &627.8  &619.7  & -          \\ 
575.5   \nii\      & -      &-      &-            &-      &-      & -           &-      &-      &-       \\ 
587.6   \hei\      & 5.8    & 5.8   &6.8          &20.9   & 20.0  & 19.7        &-      &-      &-       \\ 
654.8   \nii\      & 41.1   & 41.1  & 36.3        &-      &-      & -           &24.2   &21.7      &-        \\ 
656.3   \ha\       & 284.2  &284.2  & 280.0       &304.9  & 285.  & 281.0       &318.6  &285.   &-       \\ 
658.4   \nii\      & 109.1  &109.1  & 107.2       &1.4    & 1.3   & 1.3         &76.9   &68.7   &-       \\ 
667.8   \hei\      & -      &-      & -           &2.5    & 2.3   & 5.6        &-      &-      &-        \\ 
671.7   \sii\      & 6.0    &6.0    & 5.9         &-      &-      & -           &22.7   &20.2    &-      \\ 
673.1   \sii\      & 10.1   &10.1   & 10.1         &-      &-      & -           &17.2   &15.3    &-         \\ 
706.5   \hei\      &  -     &-      & -           &1.3    &1.2    & 1.2          &-      &-      &-      \\ 
713.5   \ariii\    & 4.6    &4.6    & 4.5         &1.     &0.9    & 0.9         &1.6    &1.     &   -       \\ 
\noalign{\smallskip} 
\cbeta            &0.0$\pm$0.02   &       &             & 0.1$\pm$0.04   &      &              &0.16$\pm$0.05   &       &        \\ 
\hline 
\noalign{\smallskip} 
\end{tabular} 
} 
\end{center} 
\end{table*} 
 
\newpage 
\begin{table*} 
\centering 
 \begin{minipage}{180mm} 
 \caption{Percentage errors in line fluxes (wrt I$_{{\rm H}\beta}$=100)} 
 \begin{tabular}{@{}llllllll@{}} 
 \hline 
Line Fluxes  & PN1 & PN2 & PN3 & PN4 & PN6 & PN7 & PN10\\ 
\hline 
1--5      &27 &30 &26 &23 &35 &25 &37 \\ 
5--15     &22 &26 &20 &17 &28 &23 &29 \\ 
15--30    &18 &22 &16 &14 &23 &15 &25 \\ 
30--200   &12 &17 &12 &10 &15 &11 &21 \\ 
200--500  &07 &11 &06 &05 &08 &06 &15 \\ 
500--1000 &05 &7  &05 &03 &05 &04 &09 \\ 
\hline 
\end{tabular} 
\end{minipage} 
\label{FluxEr} 
\end{table*} 
 
\newpage 
\begin{table*} 
\begin{minipage}{180mm} 
\caption{Physical parameters and {\sc cloudy} input parameters} 
\begin{tabular}{lllllllll} 
\hline\noalign{\smallskip} 
Parameter & \multicolumn{2}{c}{PN1} & \multicolumn{2}{c}{PN2} & \multicolumn{2}{c}{PN3} & \multicolumn{2}{c}{PN4}\\ 
                 & Empirical        &{\sc cloudy}   & Empirical     &{\sc cloudy}  & Empirical   &  {\sc cloudy}& Empirical   & {\sc cloudy} \\ 
\hline 
\te\oiii(K)                  & 13800$\pm$2600   &13300    & 15200$\pm$4000  &15100   & 15900$\pm$2700&16300   & 9200$\pm$1400 &12000  \\ 
\te\nii(K)                   & -            & -       & -       & -  & -         & -      & 14900$\pm$3000&12000 \\ 
\te\oii(K)                   & -            & -       & -       & -  & -         & -      & 16000$\pm$6000& - \\ 
\ne\sii(cm$^{-3}$)       & 4700:        & -       & -       & -  & -         & -      & 2800$\pm$600  & 3000\\ 
&&&&&&&& \\ 
\nh(cm$^{-3}$)           & -            & 5300    & -           & 5000   & -         & 5000   & -       & 2600\\ 
T$_{\star}$(K)                   & -            & 91300   & -       & 75000  & -         & 110000 & -       & 70000\\ 
$\log($L$_{\star}$/L $_{\odot}$) & -            & 2.1     & -       & 2.5    & -         & 2.5    & -       & 3.425\\ 
$\log$R$_{in}$ (cm)              & -            & 11.9    & -       & 12.    & -         & 12.    & -       & 12.\\ 
$\log$R$_{ext}$(cm)              & -            & 16.6    & -       & 16.9   & -         & 17.4   & -       & 18.\\ 
\noalign{\smallskip} 
\hline 
\end{tabular} 
\end{minipage} 
\label{Phys} 
\end{table*} 
 
\newpage 
\begin{table*} 
\begin{minipage}{140mm} 
\begin{tabular}{lllllll} 
\hline\noalign{\smallskip} 
Parameter                        & \multicolumn{2}{c}{PN6}     & \multicolumn{2}{c}{PN7}       &\multicolumn{2}{c}{PN10} \\ 
                 & Empirical     &{\sc cloudy} & Empirical   &{\sc cloudy} & Empirical   &  {\sc cloudy}\\ 
\hline 
\te\oiii(K)              & $<$14800      &12000        & 12400$\pm$3600  &12000        & -               &  -  \\ 
\ne\sii(cm$^{-3}$)               & 5000$\pm$1800 &-            & -      &-         & 100$\pm$35      &  -  \\ 
\\ 
\nh(cm$^{-3}$)                   & -         & 5800        & -           & 5000        & -           &  -  \\ 
T$_{\star}$(K)                   & -             & 84000       & -           & 71000       & -           &  -  \\ 
$\log($L$_{\star}$/L $_{\odot}$) & -             & 1.8         & -           & 3.275       & -               &  -  \\ 
$\log$R$_{in}$ (cm)              & -             & 12.         & -           & 12.         & -       &  -  \\ 
$\log$R$_{ext}$(cm)              & -             & 16.6        & -           & 17.3        & -       &  -  \\ 
 
\hline 
\end{tabular} 
\end{minipage} 
\end{table*}

\newpage 
 
\begin{table*} 
\begin{minipage}{180mm} 
\caption{{\sc icf}: ionic and total chemical abundances, and ionization correction factors} 
\begin{center} 
\renewcommand{\arraystretch}{1.4} 
\setlength\tabcolsep{5pt} 
\begin{tabular}{lllllll} 
\hline\noalign{\smallskip} 
                    & PN1       &  PN2          & PN3           &PN4            & PN6       & PN7                     \\ 
\noalign{\smallskip} 
\hline 
\noalign{\smallskip} 
He$^+$/H$^+$$\times$10$^{2}$        &8.6$\pm$1.0    & 7.5$\pm$1.0   &3.8$\pm$0.8    &14.9$\pm$2.0   &3.4$\pm$1.0   &14.0$\pm$2.     \\ 
He$^{++}$/H$^+$$\times$10$^{2}$     &$<$0.88    & 0.3$\pm$0.1   &1.2$\pm$0.5    &0.2$\pm$0.1    &0.5$\pm$0.2    &0.4$\pm$0.2            \\ 
\noalign{\smallskip} 
He/H$\times$10$^{2}$            &$<$9.44    & 7.8$\pm$1.1   &5.0$\pm$1.3    &15.1$\pm$2.1   &3.9$\pm$1.2    &14.4$\pm$2.2       \\ 
12 +log(He/H)                           &$<$10.97       & 10.9$\pm$0.10 &10.70$\pm$0.11 &11.18$\pm$0.06 &10.59$\pm$0.14 &11.16$\pm$0.07     \\ 
\noalign{\smallskip} 
CLOUDY                                  &$<$11.1        & 11.03$\pm$0.23&10.83$\pm$0.17 &11.22$\pm$0.14 &10.76$\pm$0.22 &11.27$\pm$0.15     \\ 
\noalign{\smallskip} 
\hline

\noalign{\smallskip} 
O$^+$/H$^+$$\times$10$^{7}$     &$<$2.9     & $<$4.4    &-          &57.$\pm$25.    &$<$13.7        &$<$5.0         \\ 
O$^{++}$/H$^+$$\times$10$^{5}$      &9.2$\pm$5.4    & 9.5$\pm$4.7   &7.3$\pm$3.3    &8.5$\pm$3.6    &8.3$\pm$4.3    &20.7$\pm$10.       \\ 
ICF(O)                  &1.067      & 1.028     &1.21       &1.01       &1.15       &1.03               \\ 
\noalign{\smallskip} 
 
O/H$\times$10$^{5}$         &9.8$\pm$6. & 9.8$\pm$4.9   &8.8$\pm$4.0    &9.1$\pm$3.9    &11.1$\pm$5.    &21.3$\pm$10.       \\ 
12 +log(O/H)                            &8.0$\pm$0.31   & 8.0$\pm$0.24  &7.94$\pm$0.21  &7.95$\pm$0.3    &8.04$\pm$0.21  &8.32$\pm$0.22     \\ 
\noalign{\smallskip} 
CLOUDY                                  &8.12$\pm$0.13  &8.00$\pm$0.16  &8.10$\pm$0.12  &7.98$\pm$0.14   &8.28$\pm$0.14  &8.16$\pm$0.12 \\ 
\noalign{\smallskip}

\hline 
\noalign{\smallskip} 
N$^+$/H$^+$$\times$10$^{7}$     &$<$2.3     & 6.6$\pm$3.7   &0.4$\pm$0.4    &52.$\pm$13.    &150.$\pm$50.   &2.3$\pm$2.     \\ 
ICF(N)                  &$<$342     & $<$222        &-      &15.9       &$<$8.      &$<$91.             \\ 
\noalign{\smallskip} 
N/H$\times$10$^{5}$         &$<$7.8         & $<$14.7   &-          &8.2$\pm$5. &$<$12.2        &$<$2.1     \\ 
12 +log(N/H)                            &$<$7.89        & $<$8.16       &-              &7.91$\pm$0.3   &$<$8.09        &$<$7.3     \\ 
\noalign{\smallskip} 
CLOUDY                                  &$<$7.4         &7.59$\pm 0.25$ &$<$5.7         &7.80$\pm 0.16$ &7.95$\pm 0.20$ &7.2$\pm 0.23$ \\ 
\noalign{\smallskip}

\hline 
\noalign{\smallskip} 
Ne$^{++}$/H$^+$$\times$10$^{6}$     &-          & 6.1$\pm$3.0   &1.9$\pm$1.0    &6.4$\pm$2.9    &-          &-      \\ 
ICF(Ne)                 &-      & 1.07      &1.21       &1.08       &-      &-              \\ 
\noalign{\smallskip} 
Ne/H$\times$10$^{6}$            &-          & 7.4$\pm$3.0   &2.3$\pm$1.9    &6.9$\pm$5.3    &-          &-      \\ 
12 +log(Ne/H)                           &-              & 6.87$\pm$0.20 &6.36$\pm$0.51  &6.83$\pm$0.30           &-              &-     \\ 
\noalign{\smallskip} 
CLOUDY                                  &-              & 6.91$\pm$0.21 &6.37$\pm$0.17 &6.76$\pm$0.16 &-              & - \\ 
\noalign{\smallskip} 
 
\hline 
\noalign{\smallskip} 
S$^+$/H$^+$$\times$10$^{8}$     &$<$2.8          & -            &3.2$\pm$0.1    &4.4$\pm$1.7    &16.6$\pm$ 2.   &-        \\ 
S$^{++}$/H$^+$$\times$10$^{7}$      &-           &-         &-          &-          &-          &0.3$\pm$0.2          \\ 
ICF(S)                  &4.85        & -        &-      &1.44       &2.49       &9.5           \\ 
\noalign{\smallskip} 
S/H$\times$10$^{7}$         &$<$24.      & -        &-          &6.3$\pm$2.3    &44.$\pm$8.     &11.7$\pm$3.        \\ 
12 +log(S/H)                            &$<$6.38         & -            &-              &5.80$\pm 0.14$ &6.64$\pm 0.10$ &6.06$\pm 0.11$     \\ 
\noalign{\smallskip} 
CLOUDY                                  & $<$6.56        &$<$4.86       & $<$4.28       &5.79$\pm 0.22$ &6.54$\pm 0.26$ &$<$5.98 \\ 
\noalign{\smallskip}

\hline 
\noalign{\smallskip} 
Ar$^{++}$/H$^+$$\times$10$^{7}$     &-            &0.3$\pm$0.2  &0.3$\pm$0.2    &1.0$\pm$0.3    &1.2$\pm$ 1.3   &0.7$\pm$ 0.4         \\ 
ICF(Ar)                 &-        &1.9      &1.9        &1.08       &1.87       &1.03             \\ 
\noalign{\smallskip} 
Ar/H$\times$10$^{7}$            &-            &0.6$\pm$0.5  &0.6$\pm$0.5    &1.05$\pm$0.6   &2.3$\pm$2.6    &0.7$\pm$ 0.4         \\ 
12 +log(Ar/H)                           &                 &4.78$\pm0.50$&4.78$\pm0.50$  &5.02$\pm 0.27$ &5.36$\pm 0.32$ &4.84$\pm 0.28$     \\ 
\noalign{\smallskip} 
CLOUDY                                  &-                &4.86$\pm 0.27$&4.86$\pm 0.23$&5.07$\pm 0.21$ &5.51$\pm 0.26$ &4.96$\pm 0.23$             \\ 
\noalign{\smallskip} 
 
\hline 
\end{tabular} 
\end{center} 
\end{minipage} 
\label{tab_icf} 
\end{table*} 
 
\newpage 
\begin{table*} 
\begin{minipage}{180mm} 
\caption{{\sc cloudy}: total abundances. 
The abundances are given as 12$+\log$~X/H.} 
\begin{tabular}{lllllll} 
\hline 
                &        He/H            & O/H             &N/H                     &Ne/H                 & S/H                     &Ar/H     \\ 
\hline 
PN1     &       $<$11.1          &       8.12$\pm$ 0.13    &     $<$7.4             &      -               &     $<$6.56             &     -\\ 
PN2     &       11.03$\pm$ 0.23  &       8.05$\pm$ 0.16    &     7.59$\pm$ 0.25     &     6.91$\pm$ 0.21   &     $<$4.86             &     4.86$\pm$ 0.27  \\ 
PN3     &       10.83$\pm$ 0.17  &       8.10$\pm$ 0.12    &     $<$5.7             &     6.37$\pm$ 0.17   &     $<$4.28             &     4.86$\pm$ 0.23  \\ 
PN4     &       11.22$\pm$ 0.14  &       7.98$\pm$ 0.14    &     7.80$\pm$ 0.16     &     6.76$\pm$ 0.16   &     5.79$\pm$ 0.22      &     5.07$\pm$ 0.21  \\ 
PN6     &       10.76$\pm$ 0.22  &       8.28$\pm$ 0.14    &     7.95 $\pm$ 0.20    &     -                &     6.54$\pm$ 0.26      &     5.51$\pm$ 0.26  \\ 
PN7     &       11.21$\pm$ 0.15  &       8.16$\pm$ 0.12    &     7.20 $\pm$ 0.23    &     -                &     $<$5.98             &     5.01$\pm$ 0.23  \\ 
\\ 
mean    &     11.02$^{+0.20}_{-0.36}$ & 8.06$_{-0.12}^{+0.09}$ & 7.70$_{-0.31}^{+0.18}$& 6.68$_{-0.32}^{+0.18}$& 6.35$_{-0.68}^{+0.25}$ & 5.13$_{-0.64}^{+0.24}$\\ 
\hline 
\end{tabular} 
\label{Tab_s_pn} 
\end{minipage} 
\end{table*}

\bsp 
\label{lastpage} 
\end{document}